\documentclass[aps,pra,twocolumn,superscriptaddress,showkeys,amsmath,amssymb,longbibliography]{revtex4-1}			
\usepackage{graphicx}% Include figure files
\usepackage{dcolumn}% Align table columns on decimal point
\usepackage{bm}% bold math
\usepackage[colorlinks,urlcolor=blue,citecolor=blue,linkcolor=blue]{hyperref}% add hypertext capabilities
\usepackage{subfig}
\usepackage[font=small,labelfont=bf,format=plain,justification=centerlast,labelsep=period]{caption}
\usepackage[usenames, dvipsnames]{color}

\usepackage{verbatim}
\usepackage[normalem]{ulem}

\begin{document}

\preprint{APS/123-QED}

\title{Quantum transport in non-Markovian dynamically-disordered photonic lattices}

\author{Ricardo Rom\'an-Ancheyta}
\email{ancheyta6@gmail.com}
\address{Instituto Nacional de Astrof\'isica, \'Optica y Electr\'onica, Calle Luis Enrique Erro 1, Sta. Ma.  Tonantzintla, Puebla CP 72840, Mexico}
\author{Bar{\i}\c{s} \c{C}akmak} 
\affiliation{College of Engineering and Natural Sciences, Bah\c{c}e\c{s}ehir University, Be\c{s}ikta\c{s}, Istanbul 34353, Turkey}
\author{ Roberto de J. Le\'on-Montiel}
\affiliation{Instituto de Ciencias Nucleares, Universidad Nacional Aut\'onoma de M\'exico, Apartado Postal 70-543, 04510 Cd. Mx., Mexico}
\author{Armando Perez-Leija}
\affiliation{Max-Born-Institut, Max-Born-Stra{\ss}e 2A, 12489 Berlin, Germany}
\affiliation{Humboldt-Universit\"at zu Berlin, Institut f\"ur Physik, AG Theoretische Optik {\&} Photonik, Newtonstraße 15, 12489 Berlin, Germany}
%\draft
\date{\today}

\begin{abstract}
We theoretically show that the dynamics of a driven quantum harmonic oscillator subject to non-dissipative noise is formally equivalent to the single-particle dynamics propagating through an experimentally feasible dynamically-disordered photonic network. \textcolor{black}{Using this correspondence, w}e find that noise assisted \textcolor{black}{energy} transport occurs in this network and, if the noise is Markovian \textcolor{black}{or delta-correlated, we can obtain} an analytical solution for the maximum amount of transferred energy \textcolor{black}{between all network's sites at a fixed propagation distance}. Beyond the Markovian limit, we further consider two different types of non-Markovian noise \textcolor{black}{and} show that it is possible to have efficient energy transport for larger values of the dephasing rate.
\end{abstract}

%\pacs{42.50.Pq, 32.80.−t, 42.50.Ct, 42.50.Hz}% PACS, the Physics and Astronomy
                             % Classification Scheme.
%\keywords{Suggested keywords}%Use showkeys class option if keywords

% 03.70.+k  Quantized fields
% 42.50.Pq  Quantum electrodynamics (QED) of cavities (quantum optics)
% 42.82.Et  Optical waveguides, integrated optics
% 42.50.Lc  Fluctuation phenomena quantum optics, Quantum noise
%  42.50.Hz  Relaxation processes in quantum optics
%  32.80.-t  Photon-atom interactions
%  42.50.Ct  Light interaction with matter

\maketitle

%%%%%%%%%%%%%%%%%%%%%%%%%%  body  %%%%%%%%%%%%%%%%%%%%%%%%%%

\section{Introduction}

%In general, quantum-noise processes arising from the interaction of dynamical systems with their surrounding environment is referred to as decoherence, and its understanding provides a precise explanation of the occurrence of classical behavior in nature. Simply put, decoherence is the irreversible loss of quantum coherence, or equivalently quantum superposition, which is converted into a classical incoherent mixture.

From a practical point of view, decoherence \textcolor{black}{(the irre\-versible loss of quantum coherence)} is a multifaceted process which presents advantages and disadvantages depending on the particular circumstances and application. For instance, from the perspective of \textcolor{black}{perfect} state transport, decoherence is the opponent to overcome since it may destroy the states way before they can be conveyed  \cite{SBose}. 
On the contrary, to perform highly-efficient energy transport protocols, decoherence has been found to be the best allied \cite{plenio2008,Aspuru1}. 
\textcolor{black}{Thus}, a good understanding of \textcolor{black}{the impact of decoherence on} energy transport and energy 
conversion in the quantum and classical regime\textcolor{black}{s} is 
essential to design functional technologies based on hybrid systems~\cite{Kurizki3866},
ranging from quantum information processing tasks to 
quantum thermodynamics applications ~\cite{Scully2011,Pezzutto_2019,roman2019spectral,Salado_QST_2021}.

%In quantum physics, the harmonic oscillator is one of the few analytically solvable archetypes which models a great variety of natural and synthetic systems. In this sense, in the framework of quantum optics, the quantized electromagnetic field is generally treated as a collection of harmonic oscillators \cite{Loudon}. As a result, the physics of coupled photonic systems, e.g. photonic lattices, is governed by Hamiltonians corresponding to coupled harmonic oscillators \cite{Lai1991,Bromberg2009,Tschernig2018}.

%From the opposite perspective, it is now recognized that the energy representation of a one-dimensional driven harmonic oscillator is formally equivalent to a semi-infinite lattice where the sites are represented by the amplitudes of the energy eigenstates \cite{Perez_Glauber,LeijaCorr}.
%These investigations have opened the door to using photonic waveguide systems to emulate the dynamics of the quantum harmonic oscillator in the absence of decoherence \cite{Perez_Displaced_Fock,Perez_Bloch,Wang16,ReviewSzameit2016}.
%Indeed, using integrated photonic devices, e.g. based on laser-writing coupled waveguides lattices~\cite{Szameit_2010,ReviewSzameit2016}, one can study 

\textcolor{black}{In optics, one can use integrated photonic devices,
e.g., based on direct laser-writing coupled waveguides 
lattices~\cite{Szameit_2010,ReviewSzameit2016}, to study 
coherent energy transport~\cite{LeijaPerfStateTransf,biggerstaff2015,Vinet2017coherent},
which is an important ingredient in the development 
of integrated photonic quantum technologies~\cite{OBrien2009,Sciarrino2020}.}
\textcolor{black}{Such devices constitute a well-established, popular and relatively low-cost platform among experimentalists due to their practical fabrication process, where their physical and novel geometric properties can be easily tailored~\cite{ReviewSzameit2020}. In general, these devices are never completely isolated from their environment, therefore to describe their energy losses and decoherence processes, a treatment based on the theory of open or stochastic quantum system~\cite{breuer2002theory} is needed.}

In the present work, we study coherent and
incoherent energy transport in a particular 
type of integrated photonic device \textcolor{black}{termed}
Glauber-Fock (GF) photonic lattice~\cite{Perez_Glauber,Rai_2019}. 
We choose this particular \textcolor{black}{photonic structure} because its \textcolor{black}{\it closed}
dynamics is effectively described by the unitary evolution 
of a \textcolor{black}{displaced} quantum harmonic oscillator~\cite{Perez_Glauber}, as experimentally demonstrated in ~\cite{Perez_Displaced_Fock,Perez_Bloch}.
\textcolor{black}{In addition, physics of nonlinear~\cite{Martinez_2012} and non-Hermitian systems~\cite{yuce2020diffractionfree,OZTAS20181190} can also be studied using such photonic devices.} Thus far, there is a lack of
theory accounting for the interaction of GF lattices 
with non-dissipative noisy environments. Here, we 
\textcolor{black}{close} this gap  by considering specific \textcolor{black}{instances} of
Markovian (white) and non-Markovian (correlated) noise. \textcolor{black}{Further, we show that the corresponding {\it open} system dynamics is equivalent to that of a single-excitation propagating in a dynamically-disordered network. Such n}oisy scenarios are quite relevant in integrated photonics.
For example, the interplay between noise and interference effects can lead to a faster transmission in the transport dynamics of integrated photonic mazes~\cite{caruso2015} or enhancing the coherent transport using controllable decoherence~\cite{biggerstaff2015}.
%These observations are a clear manifestation of  single-excitation environment-assisted quantum transport
\textcolor{black}{Our results also present a clear manifestation of the so-called environment-assisted transport phenomenon in the single-excitation regime~\cite{plenio2008,Aspuru1,Hauke_PRL_2019,roberto2015-2,viciani2015}. Furthermore, we observe that non-Markovianity in the dynamics of the system enhances the range of dephasing rates over which this effect persists in our model.}
%, which we consider throughout this work.} 

The paper is structured as follows. In sections~\ref{GF_section} 
and~\ref{Sing_Par_Marko} we analytically show that the master equation governing the closed and open dynamics of a \textcolor{black}{s}ingle excitation in a GF lattice is identical to the one describing the evolution of a driven quantum harmonic oscillator. We then examine the impact of non-dissipative Markovian noise on the energy transport. 
In section~\ref{Sing_Par_Non_Marko} we explore non-Markovian
noise models, while in section~\ref{results} we discuss the usefulness
of having a correspondence between the
master equations of a driven harmonic oscillator and a single particle propagating in
a GF lattice.
In particular, when the noise-assisted energy transport 
phenomenon is manifested in the Markovian case, it is possible
to find an analytical solution for the maximum amount of 
energy transferred 
between all sites of the photonic network. 
\textcolor{black}{This constitutes one of the main results of the present work and provides clear advantage in energy transport calculations.} 
For the non-Markovian case, we \textcolor{black}{find} a substantial increase in the ra\textcolor{black}{n}ge of
the dephasing rate for which the noise-assisted transport 
takes place\textcolor{black}{, indicating that the common idea~\cite{Kassal_JPC_2013} that noise-assisted transport occurs only in the moderate decoherence regime is no longer accurate when the environment's finite correlation time is considered.} In section~\ref{Conclu} we draw our conclusions.
 	
%an less attention to the non-Markovian case~\cite{chen2011excitation}

\section{The Glauber-Fock oscillator model}\label{GF_section}

We start by \textcolor{black}{introducing} the Hamiltonian of the quantum harmonic oscillator (HO) in one-dimension driven by an external  perturbation $\hat{H}=\hbar\omega  \hat{n}+\hbar g(\hat{a}+\hat{a}^\dagger)$~\cite{Castanos2019}. 
%Note that we have set the reduce Planck constant and the oscillator mass to unity, i.e., $\hbar=1$ and $m_0=1$. 
Here, $\hat{a}$ and $\hat{a}^\dagger$ are the \textcolor{black}{usual} annihilation and creation operators, $\hat{n}=\hat{a}^\dagger \hat{a}$ is the number operator, and $\omega $ is the oscillator frequency. The term $\hbar g(\hat{a}+\hat{a}^\dagger)$ represents a displacement with strength $g$. In the field of integrated photonics, this Hamiltonian governs the light dynamics in the so-called Glauber-Fock (GF) lattice~\cite{Perez_Glauber,Perez_Displaced_Fock,Perez_Bloch} or GF oscillator. 

%\subsection{Closed dynamics}\label{close_dynam}
\subsection{Unitary dynamics in GF lattices}\label{close_dynam}

Assume that the state of the driven HO is given by $|\Psi(t)\rangle=\sum_m\mathcal{A}_m(t)|m\rangle$, with $|m\rangle$ being the energy eigenstates and $\mathcal{A}_m(t)$ are the corresponding probability amplitudes. Then, it is possible to show that the equations of motion for $\mathcal{A}_m(t)$, dictated by the time-dependent Schr\"odinger equation \textcolor{black}{$i\hbar\frac{d}{dt}|\Psi(t)\rangle=\hat{H}|\Psi(t)\rangle$ with the above-mentioned Hamiltonian}, 
%with the Hamiltonian $\hat{H}=\hbar\omega\hat{n}+\hbar g(\hat{a}+\hat{a}^\dagger)$, 
are \textcolor{black}{isomorphic to} the ones describing the dynamics of \textcolor{black}{a} mode field amplitude, $\mathcal{E}_m(z)$, \textcolor{black}{propagating} in a \textcolor{black}{high-quality optical} waveguide that is coupled evanescently to its nearest neighbors forming a semi-infinite photonic lattice given as~\cite{Perez_Bloch}
\begin{equation}\label{GF_lattice_eqs}
i\frac{d}{dz}\mathcal{E}_m(z)+\mathcal{C}_m\mathcal{E}_{m-1}(z)+
\mathcal{C}_{m+1}\mathcal{E}_{m+1}(z)+\alpha m\mathcal{E}_m(z)=0,
\end{equation}
where $z$ represents the propagation coordinate,
$\mathcal{C}_m\equiv\mathcal{C}_1\sqrt{m}$ the \textcolor{black}{non-uniform} coupling 
coefficients with $\mathcal{C}_1$ 
being the coupling between the zeroth and the first waveguide
and $\alpha m$ are the propagation constants. 
%\textcolor{red}{Note that $z$ and $t$ can be interchanged by following the relation $z = \tilde{c}t$, with $\tilde{c}$ describing the effective speed of light in the medium where waveguides are inscribed}. 
Importantly, in the context of photonic lattices, the term $\alpha m$ implies that the refractive index of the waveguides describes a \textcolor{black}{potential that is gradually increasing (for $\alpha>0$) with the waveguide number. That is, the potential describes a ramp, whose slope is controlled by the parameter $\alpha$, see Fig.~1 of~\cite{Perez_Bloch} for an illustration of this refractive index profile.} \textcolor{black}{In a GF lattice $\alpha$ can also be negative, generating a different lattice response. However, throughout this work we assume that $\alpha$ is always positive.} The correspondence between Eq.~(\ref{GF_lattice_eqs}) with the equations of $\mathcal{A}_m(t)$ can be established if we identify the label 
``$m$"of each excited waveguide with the corresponding Fock 
state $|m\rangle$, $\mathcal{C}_1$ with $g$, 
$\alpha$ with $\omega$ and \textcolor{black}{the propagation coordinate} $z$ with \textcolor{black}{the time variable $t$}~\cite{ricardoguias}. 
The probability distribution, $P_m(t)=|\mathcal{A}_m(t)|^2$,
of the quantum system represents the intensity distribution, \textcolor{black}{$I_m(z)=$}
$|\mathcal{E}_m(z)|^2$, of the light in the photonic array. Details for fabricating this type of waveguide \textcolor{black}{systems} can be found in ~\cite{Perez_Bloch}.
\textcolor{black}{
For instance, to achieve the increasing coupling
$\mathcal{C}_1\sqrt{m}$ between the neighboring waveguides, 
these need to be directly inscribed in a polished fused 
silica glass using femtosecond laser writing 
technology~\cite{Szameit_2010} with a decreasing separation 
distance between them $d_m=d_1-(s/2)\ln m$, where $d_1$ and $s$ 
are parameters of $\mathcal{C}_1$ that depend on 
the corresponding waveguide width and the optical
wavelength. With this configuration the evanescent couplings 
$\mathcal{C}_m=\mathcal{C}_1\exp(-[d_m-d_1]/s)$ 
satisfy the desired square root distribution. 
}

\begin{figure}[t]
\centering
\includegraphics[width=8.65cm, height=5.5cm]{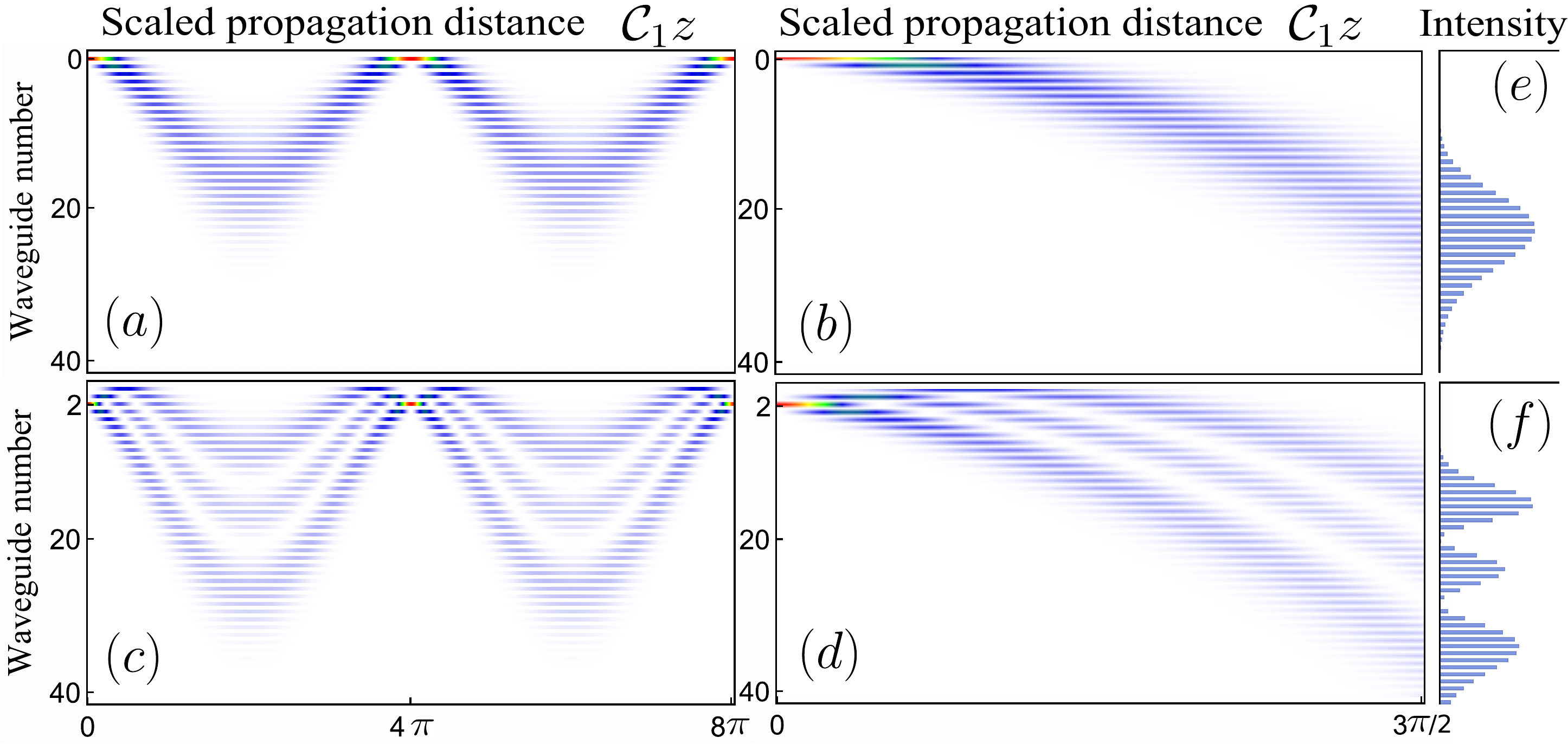} 
\caption{Light intensity propagation in a
photonic lattice obtained by integrating Eq.~(\ref{GF_lattice_eqs}).	
Light is initially launched into a single waveguide labeled
as $m=0$ {\bf (a)}, {\bf (b)} and $m=2$ {\bf (c)},{\bf (d)}.
In {\bf (e)},{\bf (f)} we \textcolor{black}{show} the output light intensity
\textcolor{black}{$I_m$}$(Z)=|\mathcal{E}_m(Z)|^2$ \textcolor{black}{at $Z=3\pi/2$ for the simulations presented in}  
{\bf (b)}, {\bf (d)}, respectively. \textcolor{black}{Note,} $Z\equiv\mathcal{C}_1z$ \textcolor{black}{is} the scaled propagation distance.
$I_m(Z)$ exhibits $m$ nodes as expected from 
to the probability distribution of the non-classical 
displaced Fock states~\cite{Perez_Displaced_Fock}.
When the ramping potential is activated, $\alpha\neq0$, we see Bloch-like oscillations~\cite{Simon_2017} 
(left), where the
light exhibits revivals every period
$Z_{\rm rev}=2\pi k \mathcal{C}_1/\alpha$ with $k$  being an 
integer ~\cite{Perez_Bloch}. 
We have set the ratio $\alpha/\mathcal{C}_1=1/2$ in {\bf (a)},{\bf (c)}
and $\alpha/\mathcal{C}_1=0$ in {\bf (b)},{\bf (d)}, which are feasible experimental values that 
\textcolor{black}{would} allow us to \textcolor{black}{implement realistic} integrated photonic devices occupying
few centimeter-scale footprints\textcolor{black}{, see the text for more details.}
%~\cite{Perez_Bloch,Perez_Displaced_Fock,Kondakci_PRA_2016,Simon_2017}.
}
\label{GF_lattices_fig}
\end{figure}

Fig.~(\ref{GF_lattices_fig}) depicts the intensity propagation 
in a GF lattice of 40 waveguides emulating the unitary 
evolution of the \textcolor{black}{driven quantum harmonic} oscillator. Clearly, \textcolor{black}{we} see 
two scenarios where the light spreads \textcolor{black}{over} the entire 
photonic \textcolor{black}{array} (delocalization)~\cite{Perez_Displaced_Fock} 
and another in which it strongly localizes as a 
manifestation of the so called Bloch-like oscillations~\cite{Simon_2017}.
\textcolor{black}{The analytical solution of Eq.~(\ref{GF_lattice_eqs}) can be found in Ref.~\cite{Perez_Bloch} in terms of the associated Laguerre polynomials. As a function of the scaled (or normalized) distance $Z\equiv\mathcal{C}_1z$, the behavior of $\mathcal{E}_m(Z)$ simply depends on the ratio $\alpha/\mathcal{C}_1$, and the so-called revival distance $Z_{\rm rev}$ is given as $Z_{\rm rev}=2\pi k\mathcal{C}_1/\alpha$. Note that the difference in the case $\alpha>\mathcal{C}_1$ as compared to $\alpha<\mathcal{C}_1$ is the period of the Bloch-like oscillations, which is short in the former case. Here, to be in accordance with the reported experimental values of $\alpha$ and $\mathcal{C}_1$ in previous works, we adopt the latter case.} 
These are simple examples showing the typical dynamics 
of coherent energy transport in closed 
systems where strong interference effects dominate.
However, when dynamical disorder mechanisms 
are considered in an open system description, a more involved study of the 
{incoherent} energy transport dynamics is necessary, which we aim to cover in the following sections.

\subsection{Open dynamics: Markovian master equation}\label{open_section}

Under the action of Markovian dephasing the density matrix $\hat{\rho}$ of the harmonic oscillator described by the Hamiltonian $\hat{H}=\hbar\omega  \hat{n}+\hbar g(\hat{a}+\hat{a}^\dagger)$  obeys the phenomenological master equation
${d\hat{\rho}}/{dt}=-i[\hat{H},\hat{\rho}]+\gamma\hat{\mathcal{L}}[\hat{n}]\hat{\rho}$.
Here the second term on the right hand side 
is the standard Lindblad superoperator given by 
$\hat{\mathcal{L}}[\hat{x}]\hat{\rho}\equiv\hat{x}\hat{\rho}\hat{x}^\dagger-\frac{1}{2}(\hat{x}^\dagger\hat{x}\hat{\rho}+\hat{\rho}\hat{x}^\dagger\hat{x})$ with $\gamma$ being the constant dephasing rate.
This master equation can be derived
using standard techniques of 
open quantum systems, where the usual Born and 
Markov approximations are used~\cite{breuer2002theory}. 

In general, pure dephasing 
processes are energy preserving~\cite{carmichael}, as a result, the interaction between the system and its environment commutes with the unperturbed Hamiltonian of the system, $\omega\hat{n}$ in the present case.
To obtain the master equation we compute the matrix elements $\langle n|\hat{H}\hat{\rho}|m\rangle=\omega  n\rho_{nm}+\sum_r gV_{nr}\rho_{rm}$, with $V_{rm}\equiv\langle r|\left(\hat{a}+\hat{a}^\dagger\right)|m\rangle=(\sqrt{m}\delta_{r m-1}+\sqrt{m+1}\delta_{r m+1})$, a similar expression is obtained for $\langle n|\hat{\rho}\hat{H}|m\rangle$.
The matrix elements for the dephasing term are
\begin{equation}\label{matx_deph}
\gamma\langle n|\hat{\mathcal{L}}[\hat{n}]\hat{\rho}|m\rangle =
\sqrt{\gamma}n\sqrt{\gamma}m\rho_{nm}-\rho_{nm}(\gamma n^2+\gamma m^2)/2.
\end{equation}
Therefore, we obtain
\begin{eqnarray}\label{matrix_element_complet_1}
i&&\frac{d}{dt}\rho_{nm}=\big[(\omega  n-\omega  m)-\frac{i}{2}(\gamma n^2+\gamma m^2)\big]\rho_{nm}\nonumber\\
&&+i\sqrt{\gamma}n\sqrt{\gamma}m\rho_{nm}
-{\sum}_r gV_{rm}\rho_{nr}+{\sum}_r gV_{nr}\rho_{rm}.
\end{eqnarray}
Defining the variables 
$\omega_n\equiv n\omega$, $\gamma_n\equiv\gamma n^2$ and $v_{ij}\equiv gV_{ij}$,
we can rewrite Eq.~(\ref{matrix_element_complet_1}) 
as
\begin{eqnarray}\label{matrix_element_complet_2}
i\frac{d}{dt}\rho_{nm}&=\Big[\big(\omega_n-\omega_m\big)-\frac{i}{2}\big(\gamma_n+\gamma_m\big)\Big]\rho_{nm}
+i\sqrt{\gamma_n\gamma_m}\rho_{nm}\nonumber\\
&\qquad-{\sum}_r v_{rm}\rho_{nr}+{\sum}_r v_{nr}\rho_{rm},
\end{eqnarray}
which is the same master equation that 
a single particle, or excitation, follows during 
its time evolution in a quantum network affected 
by non-dissipative noise, as we show in the following section (see also Eq.~(1) of Ref.~\cite{perez2018endurance}).
Since Eq.~(\ref{matrix_element_complet_2})
describes a pure-dephasing process only the off-diagonal matrix elements of 
$\hat{\rho}$ are affected by the 
constant dephasing rate $\gamma$.
%It is important to note that the form
Notice that the form
of $\gamma_n\equiv\gamma n^2$ implies that Fock states with high $n$ are more severely affected by dephasing.

\section{Single-particle dynamics in a network affected by Markovian noise}\label{Sing_Par_Marko}

In this section, we show the equivalence between 
Eq.~(\ref{matrix_element_complet_1}) 
[or \textcolor{black}{alternatively} Eq.~(\ref{matrix_element_complet_2})] and the master equation governing the evolution \textcolor{black}{(or propagation)} of a single particle
in a \textcolor{black}{tight-binding} quantum network composed of $N$ coupled sites 
affected by a stochastic non-dissipative noise (pure 
dephasing). 
\textcolor{black}{In the following and throughout the whole paper, keep in mind that with the correspondence between the spatial ($z$) and temporal ($t$) variables, the integrated photonic lattices discussed in the previous section would be a particular case of such networks.}
In order to establish this connection we begin by writing the single-particle tight-binding Hamiltonian
\begin{equation}
\hat H_S={\sum}_{n=1}^N \omega_n(t)|n\rangle\langle n|+
{\sum}_{j<n}^N\kappa_{jn}(|j\rangle\langle n|+|n\rangle\langle j|),
\label{eq:Hs}
\end{equation}
such that the evolution of the 
single-particle wavefunction, $\psi_n$, at 
the $n$th site is governed by the stochastic 
Schr\"odinger equation 
\begin{align}
\frac{d\psi_n}{dt}=-i\omega_n(t)\psi_n-i{\sum}_{j\neq n}\kappa_{nj}\psi_j,
\label{eq:SE}
\end{align}
where $\kappa_{nj}$ represents\textcolor{black}{, in principle, an arbitrary} hopping rate
between sites $n$ and $j$.
$\omega_n(t)=n\big(\omega+\phi_n(t)\big)$ is the frequency at the $n$th
site that is affected by the random fluctuations 
$\phi_n(t)$. Note each site exhibits a different natural
frequency that changes linearly with $n$, namely $\omega n$. %, i.e., $\omega_n=n\omega$. 
In most of the literature dealing with stochastic quantum networks all sites have the same frequency. However, since our main goal is to establish a connection between the present physical setting  and the one describing a driven HO, outlined in  Sec.~\ref{open_section}, we chose the frequency of the $n$th site to be proportional to $n$.
To introduce pure dephasing we consider $\phi_n(t)$ to be a Gaussian stochastic process with a
zero mean, $\langle \phi_n(t)\rangle=0$, and two-point correlation function given as
\begin{eqnarray}\label{noise_correlation}
\qquad\qquad \langle \phi_n(t)\phi_m(t')\rangle=\Gamma \delta_{nm}\delta(t-t'),
\end{eqnarray}
where $\Gamma$ is the noise strength (dephasing rate) 
that we have assumed to be the  same for all sites. 
The Kronecker delta $\delta_{nm}$ implies that the noise is uncorrelated between sites 
$n$ and $m$, the Dirac delta function 
$\delta(t-t')$ describes the Markovian nature (white noise) of
the stochastic process, and $\langle ...\rangle$ denotes the
average over all possible noise realizations. Next, following Ref.~\cite{de_J_Le_n_Montiel_2019}
we derive the corresponding master equation for the density matrix
\begin{eqnarray}\label{master_sigma_1}
i&&\frac{d}{dt}\sigma_{nm} =\big[(n\omega-m\omega)-\frac{i}{2}(\Gamma n^2+\Gamma m^2)\big]\sigma_{nm}
\nonumber\\
&&+i\sqrt{\Gamma}n\sqrt{\Gamma}m\delta_{nm}\sigma_{nm}
-{\sum}_{j }\kappa_{jm}\sigma_{nj}+{\sum}_{j }\kappa_{nj}\sigma_{jm},\quad
\end{eqnarray}
where $\sigma_{nm}(t)\equiv\langle \psi_n\psi_m^*\rangle$, see appendix~\ref{appendixA}.
Adopting the notation, $\omega_n=n\omega$ and
$\Gamma_n=\Gamma n^2$, we obtain
\begin{eqnarray}\label{matrix_element_complet_3}
i&&\frac{d}{dt}\sigma_{nm}=\Big[\big(\omega_n-\omega_m\big)-\frac{i}{2}\big(\Gamma_n+\Gamma_m\big)\Big]\sigma_{nm}
\nonumber\\ &&+i\sqrt{\Gamma_n\Gamma_m}\delta_{nm}\sigma_{nm}
-{\sum}_j \kappa_{jm}\sigma_{nj}+{\sum}_j \kappa_{nj}\sigma_{jm}.
\end{eqnarray}
Notice that the only difference between
Eq.~(\ref{matrix_element_complet_3}) and 
Eq.~(\ref{matrix_element_complet_2}) is the
Kronecker delta $\delta_{nm}$ appearing
in the second term on the right-hand-side of
Eq.~(\ref{matrix_element_complet_3}).
This difference emerges from the fact that we have 
assumed no correlation between noise affecting different 
sites, see Eq.~(\ref{noise_correlation}).
However, Eq.~(\ref{matrix_element_complet_2}) and 
Eq.~(\ref{matrix_element_complet_3}) become 
identical (no Kronecker delta in 
Eq.~(\ref{matrix_element_complet_3})) if we assume 
$\langle \phi_n(t)\phi_m(t')\rangle=\Gamma\delta(t-t')$,
i.e., there must be a correlation between 
stochastic processes at different sites.
\textcolor{black}{
Such correlation condition, which at first glance seems unlikely to achieve in practice, can easily be emulated using laser written photonic lattices in which temporal correlations are translated into longitudinal spatial correlations. In these photonic devices, ultra-short laser pulses are used to inscribe each waveguide (site) with a customized refractive index (propagation constant or site energy) depending on the writing speed~\cite{perez2018endurance}. The random fluctuations (noise) in the refractive index are implemented by modulating the laser's writing speed during the manufacturing process with a high degree of control, keeping the coupling coefficients unchanged~\cite{caruso2015}.}
\textcolor{black}{Contrary to uncorrelated noise~(\ref{noise_correlation}), where independent noise generators in each inscribed waveguide are used~\cite{caruso2015,perez2018endurance}, for correlated noise between sites ($\delta_{nm}=1$), a single generator would need to be used during each fabrication step of the waveguide array.}
%
%Contrary to uncorrelated noise~(\ref{noise_correlation}), where independentnoise generators in each waveguide are used [22, 30], forspatially-correlated noise (12), a single generator wouldneed to be used in each propagation step of the waveguidearray. 

\textcolor{black}{Let us emphasize that in GF lattices, the hopping rates must satisfy, as in the previous case, $\kappa_{nj}=g(\sqrt{j}\delta_{n j-1}+\sqrt{j+1}\delta_{n j+1})$ and the time evolution must be interpreted as spatial propagation. The light intensity represents the probability distribution $P_n$ but now this is given by the diagonal matrix elements $\rho_{nn}$ and $\sigma_{nn}$.}

\begin{figure}[t]
\centering
\includegraphics[width=8.65cm, height=5.4cm]{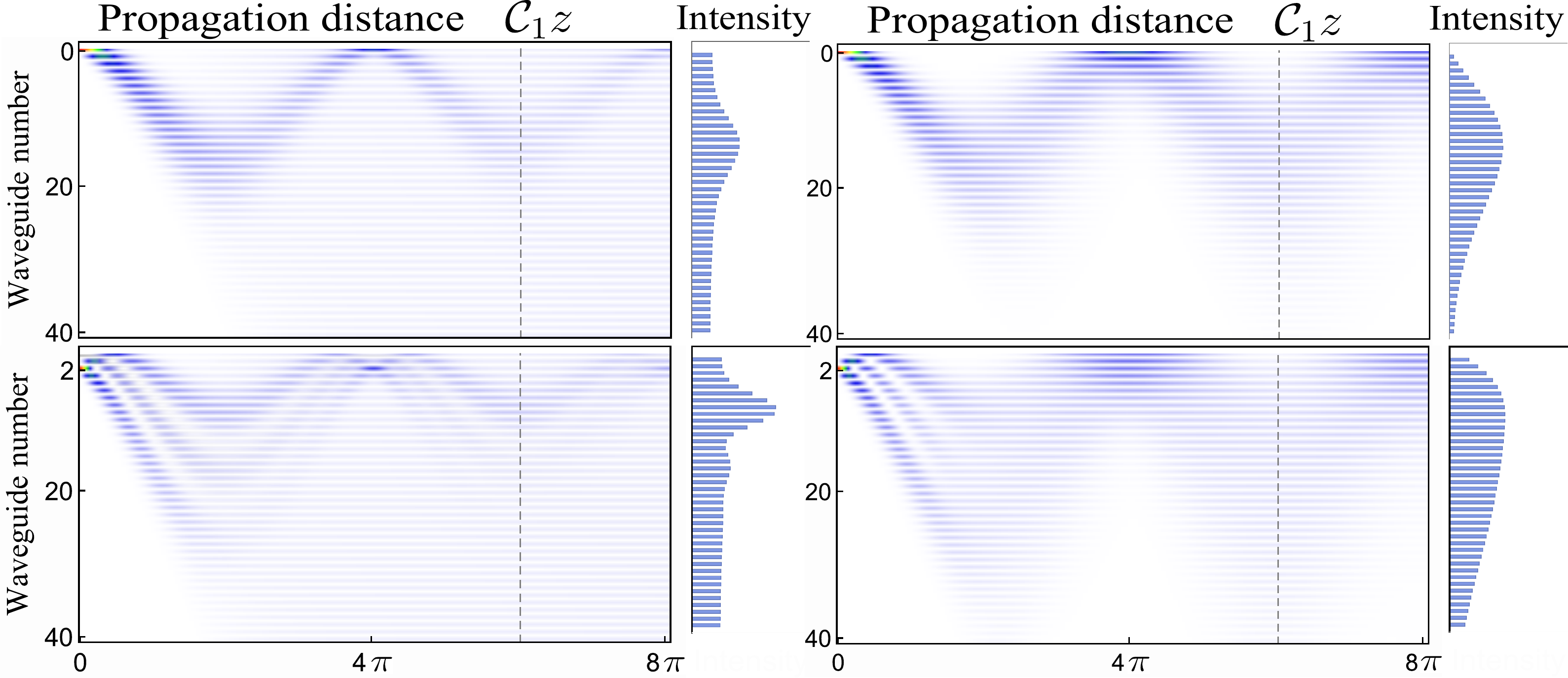} 
\caption{Light propagation in a dynamically disorder Glauber-Fock lattice. On the left we see the diagonal elements 
$\sigma_{nn}$ obtained by numerical integration of Eq.~(\ref{matrix_element_complet_3}) describing
the emerging intensity distribution in the lattice, on the right we depict the matrix elements
$\rho_{nn}$ from Eq.~(\ref{matrix_element_complet_2}). For these simulations we used the same parameters 
and initial conditions as for Figs.~(\ref{GF_lattices_fig}~$a$,$c$) 
but now adding white noise with a dephasing rate $\Gamma=0.001$ (left)
and $\gamma=0.025$ (right). Due to the noise, light 
starts to delocalize and the effect is more prominent in the 
regions where complete Bloch-like oscillations would appear 
in the absence of noise, see Fig.~(\ref{GF_lattices_fig}).
Vertical dashed line indicates the distance at which the light 
intensity is calculated.}
\label{GF_lattices_fig_noise}
\end{figure}

In Fig.~(\ref{GF_lattices_fig_noise}) we compare the dynamics 
generated by numerically integrating Eq.~(\ref{matrix_element_complet_3})
and Eq.~(\ref{matrix_element_complet_2}).
Specifically, we present the 
corresponding diagonal elements, $\sigma_{nn}$ 
and $\rho_{nn}$, to illustrate the intensity propagation in 
a dynamically disorder 
Glauber-Fock photonic lattice. 
Both master equations were numerically solved
using the technique described in ~\cite{navarrete2015open}.  
\textcolor{black}{In Fig.~(\ref{GF_lattices_fig_noise}) one can see that due to the added noise, near the revival  distances $\mathcal{C}_1z_{\rm rev}=4\pi k$ light delocalization is more prominent. From the experimental point of view this means that one could build photonic waveguide arrays having just one Bloch-like oscillation ($k=1$) and see the desired dephasing effect. In fact, the ratio $\alpha/\mathcal{C}_1=1/2$ used in Figs.~(\ref{GF_lattices_fig}) and~(\ref{GF_lattices_fig_noise}) can easily be obtained by choosing the coupling between the zeroth and the first waveguide as $\mathcal{C}_1=0.88{\rm cm}^{-1}$~\cite{Kondakci_PRA_2016,Martin_OE_2011} and $\alpha=0.044{\rm mm}^{-1}$~\cite{Simon_2017}. The use of these values implies that we should design approximately $40$ nearest-neighbor coupled waveguides with $z_{\rm rev}=4\pi/\mathcal{C}_1=14.28{\rm cm}$, which is a feasible scenario. For instance, in~\cite{LeijaPerfStateTransf,biggerstaff2015,Perez_Displaced_Fock,Perez_Bloch,Stutzer_2013} waveguides arrays of $10{\rm cm}$ to $15{\rm cm}$ long were built, and in~\cite{Kondakci_PRA_2016,Martin_OE_2011}, $101$ identical waveguides were inscribed within one of these types of arrays. In the photonic device, the parameter $\mathcal{C}_m=\mathcal{C}_1\sqrt{m}$ increases with the waveguide's label, and by using the above parameters we obtain $\mathcal{C}_{40}\approx 5.5{\rm cm}^{-1}$, which corresponds to the largest coupling coefficient reported experimentally in~\cite{Bloch_Zener_PRL_2009}.}
\textcolor{black}{It is worth pointing out that, given the stochastic nature of the process, a certain number of waveguide samples is needed in order to observe the mean density matrix described in Eq.~(\ref{matrix_element_complet_3}). Previous work by two of us~\cite{perez2018endurance} has shown that this number is approximately $20$.} 
\textcolor{black}{We would also like to mention that reconfigurable electrical oscillator networks~\cite{roberto2015-2,roberto_PRR_2021} and optical tweezer arrays \cite{quinto2017,quinto2019} are other viable experimental platforms in which Eq.~(\ref{matrix_element_complet_3}) and the correlated noise condition between different sites can be realized.}

\section{Time-dependent dephasing rate in the master equation}\label{Sing_Par_Non_Marko}

We now turn our attention to generalize the results obtained in the previous section to the case of time-dependent dephasing.
This opens up the possibility to introduce memory effects in the dynamics of both models, that is, it enables investigations of non-Markovian effects.

\subsection{Glauber-Fock oscillator}
The master equation, in the Lindblad form, for the Hamiltonian
$\hat{H}$ under a time-dependent dephasing 
noise is given as
\begin{equation}\label{METD}
\frac{d\hat{\rho}}{dt}=-i[\hat{H},\hat{\rho}]+\gamma(t)\hat{\mathcal{L}}[\hat{n}]\hat{\rho}.
\end{equation}
This equation is an ad hoc generalization of 
the master equation of a two-level system 
describing pure-dephasing dynamics 
in a possibly non-Markovian regime (see Eq.~(9) of 
Ref.~\cite{YU2010676} and Eq.~(17) of 
Ref.~\cite{Maniscalco2014}).
In cases where 
$\gamma(t)$ becomes negative, 
the quantum dynamical semigroup property of 
Eq.~(\ref{METD}) no longer
holds~\cite{breuer2002theory}. Consequently, 
the divisibility of the quantum map is 
broken and Eq.~(\ref{METD})
can be classified as non-Markovian~\cite{Maniscalco2014,PRA_Hall}. 
Here, we only consider cases in which $\gamma(t)$ 
is non-negative. 
However, it is worth \textcolor{black}{pointing out} that Eq.~(\ref{METD}) 
can be used to describe the dynamics of the system under non-Markovian environments provided the dephasing rates exhibit finite environment correlation times ~\cite{YU2010676,kumar2018non}.\\
From Eq.~(\ref{METD}), and following the same steps as in the previous section, we obtain the master equation 
\begin{eqnarray}\label{matrix_elements_time}
&&i\frac{d}{dt}\rho_{nm}=\Big[\big(\omega_n-\omega_m\big)-\frac{i}{2}\big(\gamma_n(t)+\gamma_m(t)\big)\Big]\rho_{nm}
\nonumber\\ &&+i\sqrt{\gamma_n(t)\gamma_m(t)}\rho_{nm}
-{\sum}_r \kappa_{rm}\rho_{nr}+{\sum}_r \kappa_{nr}\rho_{rm}.\quad
\end{eqnarray}
\textcolor{black}{It is interesting to note that the} only difference between this expression
and Eq.~(\ref{matrix_element_complet_2}) 
is the time-independent $\gamma_n$ which is now replaced by 
$\gamma_n(t)\equiv\gamma(t)n^2$. In what follows, we discuss its effect in the transport dynamics of complex quantum networks.

%\newpage
\subsection{Single-particle dynamics in a quantum network}
We start by investigating the dynamics of a single particle under the influence of a Gaussian 
non-Markovian stochastic noise
$\Omega_n(t)$, with zero mean 
$\langle \Omega_n(t)\rangle=0$ and two-point correlation function
\begin{equation}\label{noise_correlation_memory}
2\left\langle \Omega_n(t)\Omega_m(t')\right\rangle={\Gamma\lambda}\delta_{nm}e^{-\lambda|t-t'|}.
\end{equation}
This is the well known modified Ornstein-Uhlenbeck 
noise (OUN), where $\Gamma$ is the inverse relaxation time and $\lambda$ is the noise bandwidth which is related to the environmental memory time in the following way $\tau_c=\lambda^{-1}$, such that when $\lambda$ is finite the $\tau_c$ is also finite giving a non-Markovian character to the dynamics (see Eq.~(3) of~\cite{YU2010676} 
and Eq.~(2.9) of Ref.~\cite{kumar2018non} for a detailed discussion on OUN). This process 
has a well defined Markovian limit which is obtained when
$\lim_{\lambda\rightarrow\infty} \langle \Omega_n(t)\Omega_m(t')\rangle=\Gamma\delta_{nm}\delta(t-t')$~\cite{YU2010676} .\\
As shown in appendix~\ref{appendixA}, for the present case we obtain the equation
\begin{eqnarray}\label{rhonm_non_first}
&&\frac{d}{dt}\sigma_{nm}=-i(n\omega-m\omega)\sigma_{nm}
-i{\sum}_{j\neq n}\kappa_{nj}\sigma_{jm}
\nonumber\\&&+i{\sum}_{j\neq m}\kappa_{jm}\sigma_{nj}
-in\langle\psi_n\psi_m^*\Omega_n(t)\rangle
+im\langle\psi_n\psi_m^*\Omega_m(t)\rangle,\nonumber\\
\end{eqnarray}
where the matrix elements are given as $\sigma_{nm}=\langle \psi_n\psi_m^*\rangle$.\\
Although $\Omega_n(t)$ represents in general 
non-Markovian noise it is  still a Gaussian 
process, therefore, we can use the Novikov's 
theorem~\cite{novikov1965} to compute the elements
\begin{align}\label{Psi_nPsi_mOmegt}
\langle\psi_n\psi_m^*\Omega_n(t)\rangle\approx -\frac{i}{2}n\sigma_{nm}(t)\Gamma(t)+\frac{i}{2}m\delta_{mn}\sigma_{nm}(t)\Gamma(t),
\end{align}
where $\Gamma(t)\equiv{\Gamma}(1-e^{-\lambda t})/2$, 
is, basically, the time integral of the two-point 
correlation function of the environment~\cite{YU2010676},
see appendix~\ref{appendixB} for details. 
Hence, Eq.~(\ref{rhonm_non_first}) becomes
\begin{eqnarray}\label{matrix_elements_time_non}
&&i\frac{d}{dt}\sigma_{nm}=\Big[\big(\omega_n-\omega_m\big)-\frac{i}{2}\big(\Gamma_n(t)
+\Gamma_m(t)\big)\Big]\sigma_{nm}
\nonumber\\&&+i\sqrt{\Gamma_n(t)\Gamma_m(t)}\delta_{nm}\sigma_{nm}
-{\sum}_r \kappa_{rm}\sigma_{nr}+{\sum}_r \kappa_{nr}\sigma_{rm},
\nonumber\\
\end{eqnarray}
where $\Gamma_n(t)\equiv\Gamma(t)n^2$. 
This is the master equation and describes the 
dynamics of a single particle in a 
non-Markovian environment and, as expected, it
reduces to Eq.~(\ref{matrix_element_complet_3})
when $\Gamma(t)$ is time-independent. 

Similarly to the case of Markovian noise, discused in the previous section, the only difference 
between Eq.~(\ref{matrix_elements_time_non})
and Eq.~(\ref{matrix_elements_time}) is the 
additional Kronecker delta appearing in
Eq.~(\ref{matrix_elements_time_non}).
That is, when there are noise correlations between different sites
then $\rho_{nm}$ 
and $\sigma_{nm}$ become identical. Even though they exhibit similar evolution, 
$\rho_{nm}$ and $\sigma_{nm}$ are of a
different nature. In other words, $\rho_{nm}$
is the density matrix for a quantum harmonic
oscillator inhabiting in an infinite dimensional Hilbert space spanned by 
an infinite number of Fock states. 
In contrast, $\sigma_{nm}$ is the density matrix of a single particle, 
or excitation, evolving in a quantum network
made out of finite number of coupled sites, with their 
corresponding Hilbert space. 

\textcolor{black}{Finally, we would like to stress that the derivation of Eqs.~(\ref{matrix_element_complet_3}) and (\ref{matrix_elements_time_non}) is, in principle, valid for arbitrary time-independent hopping rates $\kappa_{rm}$ and not only for nearest-neighbor interactions. Therefore, these master equations may describe an extensive class of complex networks that do not necessarily have to be photonic.}

\section{Average energy of the system}\label{results}

\subsection{Analytical solution for the Ma\textcolor{black}{r}kovian case}

\begin{figure}[t]
\centering
\includegraphics[width=7cm, height=5cm]{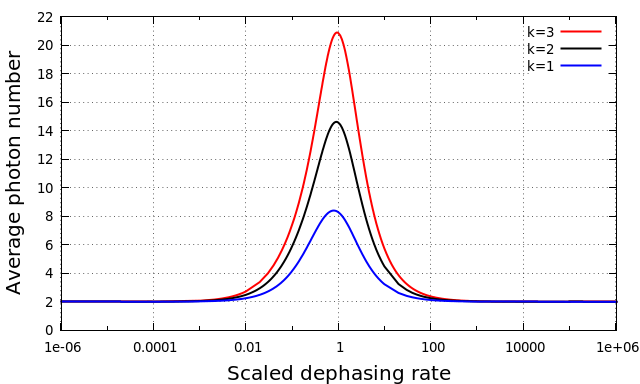} 
\caption{Manifestation of the noise assisted transport 
phenomenon, captured by the average photon number of 
Eq.~(\ref{eq:nois:assi}), as a function of the scaled 
dephasing rate $\tilde{\gamma}=\gamma/\omega $ at 
different scaled revivals times counted by $k$, ratio $g/\omega=1$ and initial condition $m=2$.}
%The system is initially in the state $|2\rangle$.}
\label{better_1}
\end{figure}

In this section we discuss some advantages of having 
a correspondence between the master equations of a driven \textcolor{black}{quantum} harmonic 
oscillator and a single particle propagating in a \textcolor{black}{photonic} network.
When one is interested in computing the average of certain observables, e.g. the average energy of a particle propagating in a network, solving the HO master equation is much simpler than solving the latter.
For example, if we insert the Hamiltonian describing the Glauber-Fock
oscillator
%
%\begin{align}\label{Hx}
$\hat{H}_{\texttt{GF}}\equiv \omega \hat{n}+g(\hat{a}+\hat{a}^\dagger)$
%\end{align}
%
into Eq.~(\ref{METD}),
we readily obtain the equations of 
motion for the average of the number and field operators
\begin{eqnarray}\label{Hx}
\frac{d\langle \hat n\rangle}{dt}=-i g\langle \hat a^\dagger\rangle+ig\langle \hat a\rangle,\,\,
\frac{d\langle \hat a\rangle}{dt}=\left(-i\omega -\gamma(t)\right)\langle \hat a\rangle -ig,\nonumber\\
%\frac{d\langle\hat a^\dagger\rangle}{dt}=\frac{\langle \hat a \rangle^*}{dt}.
\end{eqnarray}
where ${d\langle\hat a^\dagger\rangle}/{dt}={d\langle \hat a \rangle^*}/{dt}$.
In the absence of dephasing, $\gamma(t)=0$, and assuming the initial condition $|\psi(0)\rangle=|m\rangle$, these equations reduce to the well-known solution for the average of the number operator
$\langle \hat n(t)\rangle=m+({2g}/{\omega })^2 \sin^2(\omega  t/2)$. 
From this expression, one can directly determine the time at which the states return to its initial configuration (revival time) 
$t_\mathtt{rev}=2\pi k/\omega $, 
with $k$ being an integer. 
Further, in the limit $\omega\rightarrow 0$, the evolution operator becomes the Glauber displacement displacement operator, $D(\alpha)=\exp(\alpha\hat a-\alpha^*a^\dagger)$ with $\alpha=igt$, that transforms a vacuum initial state into a coherent state~\cite{Perez_Displaced_Fock}. Accordingly, in this limit we have $\lim_{\omega \rightarrow 0}\langle \hat n(t)\rangle\sim(gt)^2$,
which corresponds to the average value of a coherent state.
For the case of non-dissipative (pure dephasing) Markovian noise the dephasing rate is a non-negative constant $\gamma\neq 0$. 
Then, the average for the number operator is
\begin{eqnarray}
\langle \hat n(t)\rangle&=&m+\frac{2g^2}{(\omega^2+\gamma^2)^2}\big(f(t)+e^{-\gamma t}g(t)\big),%\\
\end{eqnarray}
where we have defined the functions
$f(t)\equiv\gamma^2(\gamma t-1)+\omega^2(\gamma t+1)$
and
$g(t)\equiv(\gamma^2-\omega^2)\cos(\omega  t)-2\gamma\omega \sin(\omega  t)$.
It is remarkable that \textcolor{black}{the} temporal behavior of $\langle\hat{n}(t)\rangle$ in the quantum system
gives crucial information about the energy transport across all sites in the photonic structure at fixed propagation distance~\cite{ricardoguias}. This is because $\langle\hat{n}(t)\rangle$ can be rewritten as $\langle\hat{n}(t)\rangle=\sum_nP_n(t)n$, where $P_n(t)=\rho_{nn}(t)$ is the probability distribution that we are associating with the light intensity $I_n(z)$ on each waveguide of the photonic lattice. 
So, in GF lattices, $I_n(z)$ is measured first and then the quantitiy $\langle \hat{n}(z)\rangle_{\rm class}\equiv\sum_{m=0}^N m I_m(z)$ is evaluated. This corresponds to the classical analog of the average photon number in waveguide arrays. 
Evaluating $\langle\hat{n}(t)\rangle$ at the revival time \textcolor{black}{(revival distance in the GF lattice)} yields
\begin{eqnarray}\label{eq:nois:assi}
\langle \hat n(t_\mathtt{rev})\rangle=
m+\frac{2g^2}{\omega^2}\Big[ 
\frac{2\pi k \tilde\gamma}{1+\tilde\gamma^2}+
\frac{1-\tilde\gamma^2}{(1+\tilde\gamma^2)^2}\big(1-e^{-2\pi k \tilde\gamma}\big)\Big],
\nonumber\\
\end{eqnarray}
where $\tilde{\gamma}=\gamma/\omega $ is the scaled
dephasing rate and $m$ is the initially excited site.
Note $\langle \hat n(t_\mathtt{rev})\rangle$
attains its maximum value when the decoherence rate is comparable 
to the energy scale of the system, i.e., when $\tilde{\gamma}\sim 1$ 
Eq.~(\ref{eq:nois:assi}) reduces to
$\langle \hat n(t_{\rm rev})\rangle_\mathtt{max}\sim m+2\pi k\left({g}/{\omega }\right)^2$. 
%These results are similar to the ones reported in Ref.~\cite{Aspuru1}.
This result indicates that independently of the initial
condition \textcolor{black}{(excited site)} $|m\rangle$, the delocalization of the initial 
excitation will increase 
linearly, as a function of $k$, at each revival time \textcolor{black}{(distance)} as 
shown in Fig.~(\ref{better_1}).

The Bell-like shape depicted in Fig.~(\ref{better_1}) is in fact a signature of the \textcolor{black}{so-called} environment assisted transport  phenomenon, \textcolor{black}{which} in the present case is symmetric with respect to the scaled dephasing rate. \textcolor{black}{Note that this result contrasts} with the asymmetric behavior typically observed in other \textcolor{black}{coupled-oscillator} systems~\cite{ricardoguias,Hauke_PRL_2019,Robert_PRL}.\\
%(see Fig.~4.(a) in~\cite{ricardoguias}, Fig.~1 in~\cite{Hauke_PRL_2019} and Fig.~2 in~\cite{Robert_PRL}).\\
%
In general, there are two distinct regimes in systems exhibiting noise-assisted transport.  
For small dephasing rate, $\tilde{\gamma}\ll 1$, 
the energy transport is proportional to 
$\tilde{\gamma}$, such that Eq.~(\ref{eq:nois:assi}) reduces to
$\langle \hat{n}(t_{\rm rev})\rangle\sim
m+4\pi k({g}/{\omega})^2\tilde{\gamma}$.
On the other hand, when the dephasing rate is very high, 
$\tilde{\gamma}\gg1$, the energy transport decreases
with $1/\tilde{\gamma}$, in fact, it is easy to show that 
$\langle \hat{n}(t_{\rm rev})\rangle\sim m+4\pi k({g}/{\omega})^2{\tilde{\gamma}}^{-1}$.
These regimes are shown in
Fig.~(\ref{fig2}), see black solid lines.
The fact that the energy transport has a nonmonotonic 
behavior can be understood, in a quantum scenario, as 
a consequence of the quantum Zeno effect (QZE)
~\cite{SebastianDiehl2019}. In the QZE a frequent 
measurement on a quantum system inhibits transitions 
between quantum states~\cite{Sudarshan1977}. In our system the QZE is 
dominant when the dephasing rate is extremely high,
i.e, when the non-disipative noise acts as the measurement
process.

\textcolor{black}{In the corresponding optical context of waveguide arrays, the above effects are expected to occur at the revival distance, under the assumptions of couplings coefficients without disorder and no losses in the waveguide array. However, that is not the case in practical implementations. For example, in the presence of static disorder Anderson localization of light will occur for large disorder values as experimentally demonstrated in~\cite{Martin_OE_2011} for $\alpha=0$, i.e., without Bloch oscillations. When static disorder and Bloch oscillations are both present, Hybrid Bloch-Anderson localization of light emerges with gradual washing out of Bloch oscillations~\cite{Stutzer_2013}. Nevertheless, in such case, the first Bloch-like revival (the one we have required to be present in this work) is still visible~\cite{Stutzer_2013}. Hence, we deduce that our results are robust against static disorder.}
\textcolor{black}{On the other hand, a typical experiment of this kind shows low losses. To be more specific, propagation losses are in the range of 0.1$-$0.9${\rm dB~cm}^{-1}$ for straight sections of the waveguides and also there is an excellent mode overlap with standard fibers ($0.1\,{\rm dB~cm}^{-1}$)~\cite{ReviewSzameit2016,ReviewSzameit2020,perez2018endurance}. Moreover, losses are approximately independent of the writing speed~\cite{Szameit_2010}.}

\textcolor{black}{Before concluding this subsection, we would like to point out that while we have assumed zero temperature conditions for the driven quantum harmonic oscillator, there is no restriction for considering temperature effects upon the photonic lattices structures. Most experiments using direct laser-written waveguides are performed at room temperature~\cite{Szameit_2010,ReviewSzameit2016}. Moreover, impressive thermal effects can be admitted on these devices. For instance, in~\cite{Pertsch_PRL_1999}, it was experimentally shown that varying a temperature gradient, the Bloch oscillations' period and amplitude could be controlled. This can be done by heating and cooling the opposite sides of the waveguide array. Specifically, a transverse linear temperature gradient $\Delta T$ leads to a linear variation of the propagation constants, i.e., $\alpha\propto\Delta T$. Such result suggests an attractive alternative to get the desired ramp potential or the random fluctuations without changing the laser's writing speed.}

\subsection{Numerical solution for the non-Markovian case}

We now look into the numerical solutions of the equations of motion for the average field and number operators
under two different types of non-Markovian noise models. We consider Ornstein-Uhlenbeck noise (OUN) and power law noise (PLN), both of which have a well defined Markovian limit. The time-dependent dephasing rates for these cases are given as~\cite{kumar2018non,Shrikant2019}
\begin{equation}
  \Gamma(t) =
    \begin{cases}
      \frac{\Gamma}{2}(1-e^{-\lambda t}) & \text{OUN},\\
      \frac{\Gamma}{2(\lambda t+1)^3} & \text{PLN},
    \end{cases}       
\end{equation}
where $\Gamma$ is the inverse relaxation time and $\lambda$ and $1/\lambda$ are the noise bandwidths for OUN and PLN, respectively, which in turn are related to the finite correlation time of the environment. Note these quantities can be considered as the inverse environmental memory time that vanish in the limit of $\lambda\rightarrow\infty$ and $1/\lambda\rightarrow\infty$ for OUN and PLN, respectively, yielding the Markovian limits of these noise models. Naturally, in the Markovian limit, the time-dependent dephasing factors become time-independent.
%, that is, we have $\Gamma/4$ and $\Gamma/2$ for OUN and PLN, respectively. 
It is important to note that in both cases $\Gamma(t)$ never becomes negative throughout the evolution. Therefore, the dynamics generated are CP-divisible at all times and considered as Markovian~\cite{Rivas2014,RMP_Breuer}. Nevertheless, it is clear that finite environment correlation time results in non-Markovian behavior in the dynamics~\cite{YU2010676,kumar2018non}, such that any intermediate map taking the system from $t_1$ to $t_2$, is not independent of the initial time $t_0$. It has been shown that it is also possible to quantify these ``weaker" forms of non-Markovianity emerging in these models, by adopting different strategies as shown in~\cite{Shrikant2019}.
\begin{figure}[t]
%{\bf (a)} \hskip8cm {\bf (b)}\\
%{\hspace{2.5cm}
\bf (a)\\
\includegraphics[width=0.8\columnwidth]{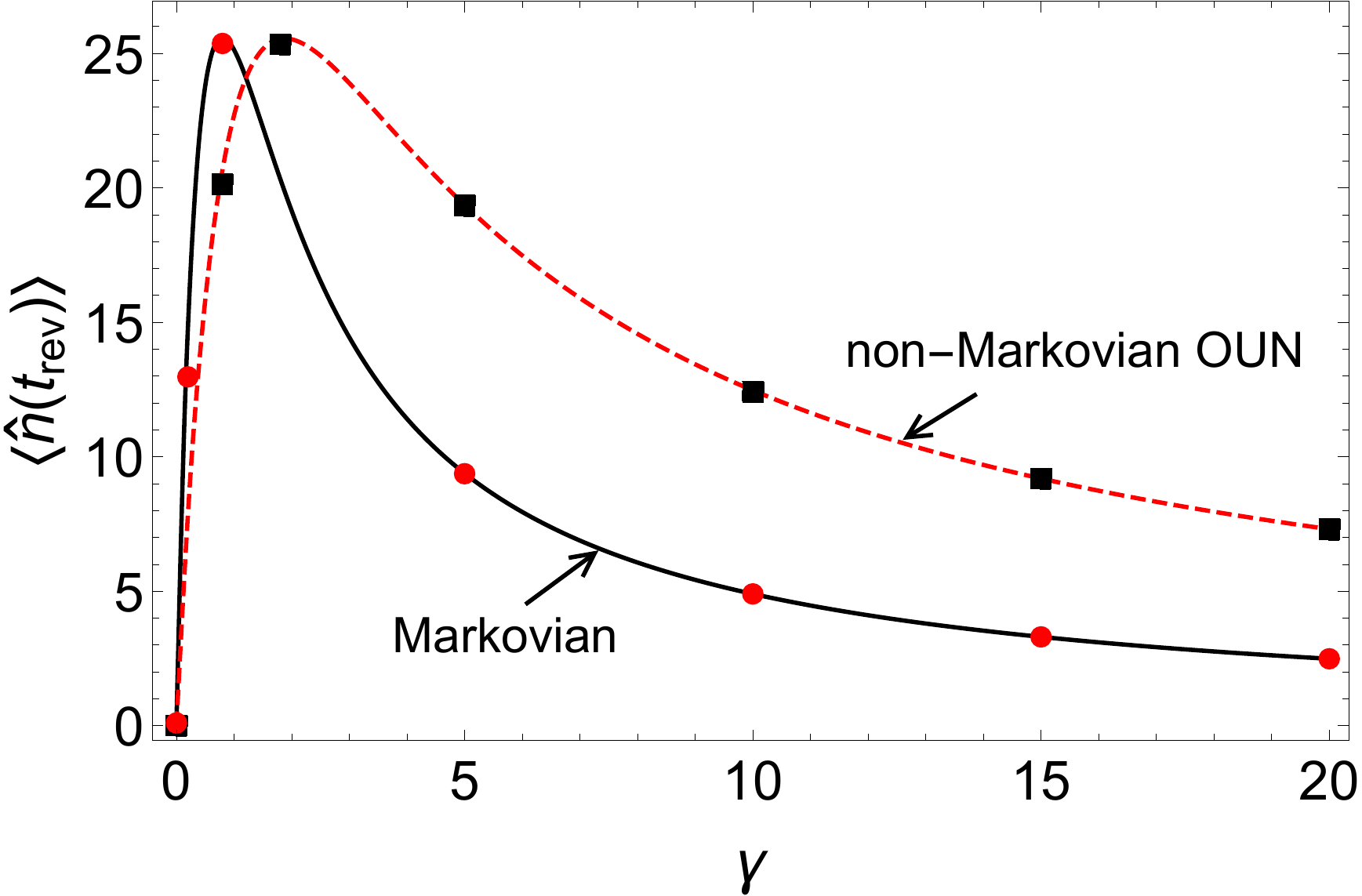}\\
%{\hspace{-0.5cm}
\bf (b)\\
\includegraphics[width=0.8\columnwidth]{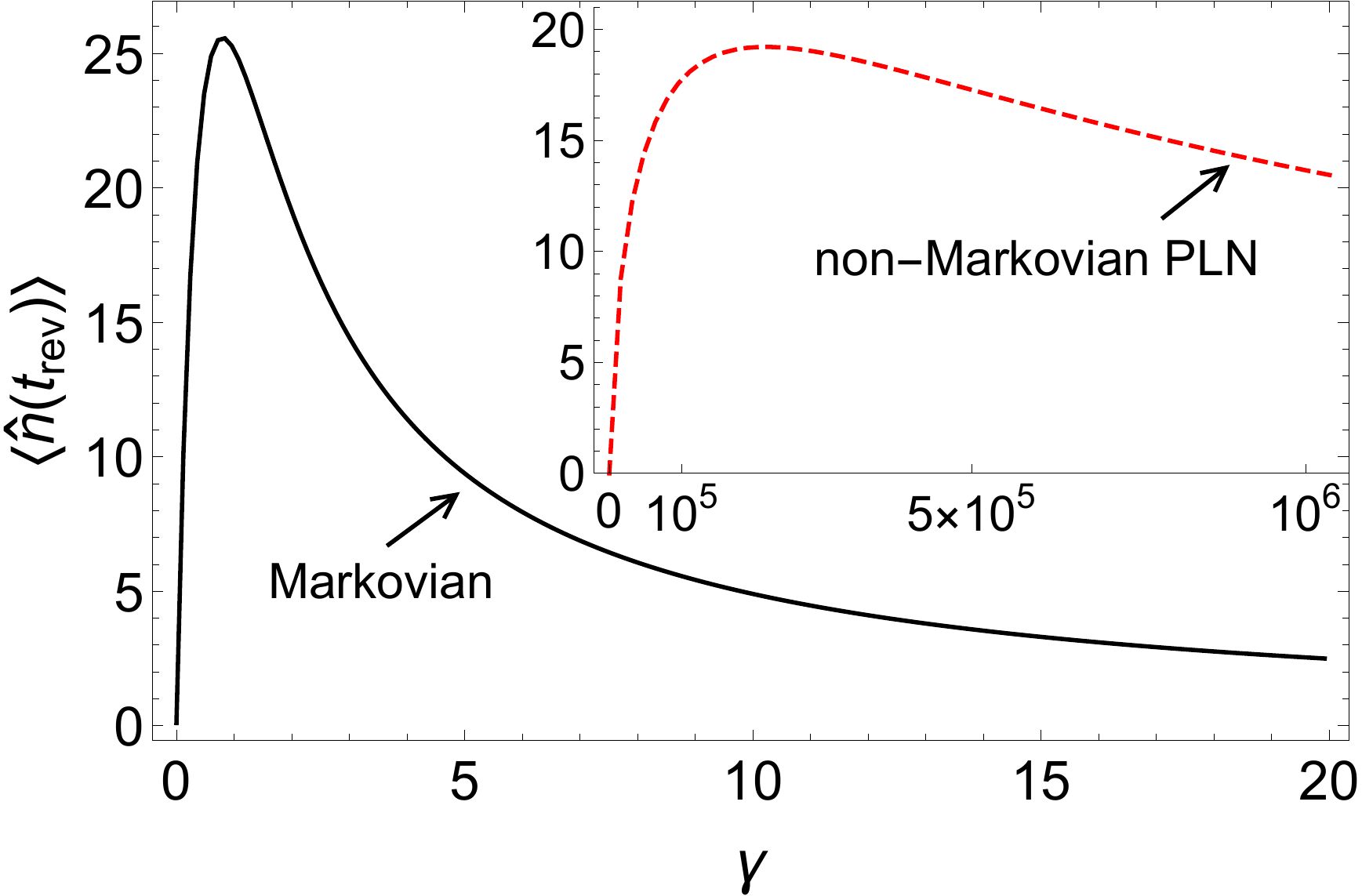}\\
\caption{Noise-assisted transport phenomenon in a Glauber-Fock oscillator \textcolor{black}{(lattice)} as a function of the dephasing rate, $\gamma$, under \textbf{(a)} \textcolor{black}{Ornstein-Uhlenbeck noise} (OUN) and \textbf{(b)} \textcolor{black}{power law noise} (PLN). The system was prepared initially in ground state $|0\rangle$, i.e., $m=0$, $g=1$, $\omega=0.5$ and $k=1$. For \textbf{(a)} (\textbf{(b)}) $\lambda\rightarrow\infty$ ($\lambda^{-1}\rightarrow\infty$) black solid line corresponds to the Markovian limit \textcolor{black}{given by  Eq.~(\ref{eq:nois:assi}),} and $\lambda^{-1}=\textcolor{black}{10}$ ($\lambda= \textcolor{black}{10}$) red dashed line shows the behavior in the non-Markovian regime \textcolor{black}{evaluating Eq.~(\ref{Hx}) at $t_{\rm rev}$. We obtain the red (black) circular (square) dots by integrating the master equation~(\ref{matrix_elements_time}).}}
\label{fig2}
\end{figure}

In~\cite{Shrikant2019}, the authors \textcolor{black}{provide} a geometric measure of non-Markovianity that is capable of capturing the amount of non-Markovianity for the CP-divisible models considered above and provide a comparative analysis. Fixing $x=\lambda^{-1}$ for OUN and $x=\lambda$ for PLN, it has been shown that the non-Markovianity of PLN is always higher than that of OUN for any finite $x$ (cf. Fig. 1-(a) of ~\cite{Shrikant2019}). Note that $x=0$ corresponds to the case $\lambda\rightarrow\infty$ for OUN and $1/\lambda\rightarrow\infty$ for PLN, which are the Markovian limits of these models. We set $x=\textcolor{black}{10}$ and look at the noise-assisted transport phenomenon in a Glauber-Fock oscillator  \textcolor{black}{(lattice)} under OUN and PLN noise, together with their corresponding Markovian limit.
Fig.~(\ref{fig2}) shows that for this value, $x=\textcolor{black}{10}$, which corresponds to $\lambda=\textcolor{black}{0.1}$ for OUN and $\lambda=\textcolor{black}{10}$ for PLN, the noise-assisted transport phenomenon \textcolor{black}{shows a higher enhancement over a broader range of dephasing} in the case of PLN as compared to OUN, which also presents a higher non-Markovianity. 

\textcolor{black}{In Fig.~\ref{fig2}$(a)$, we show that the pure numerical calculation of the master equation~(\ref{matrix_elements_time}) coincides (as expected) with the solution of Eq.~(\ref{Hx}), but the latter is significantly more straightforward to solve than the former. Interestingly, even when we do not consider correlations between sites, one can still observe the enhancement in noise-assisted transport in the non-Markovian case described by the master equation~(\ref{matrix_elements_time_non}).} 
\textcolor{black}{This suggests that for the specific model of GF oscillator (lattice) considered in this work, non-Markovianity seems quite advantageous in the noise-assisted transport phenomenon. Increased non-Markovianity in the open system dynamics, allows us to achieve finite $\langle \hat n(t)\rangle$,  or equivalently $\langle\hat{n}(z) \rangle_{\rm class}$, for larger values of the dephasing rate. Although our findings are limited to the models considered in this work, they are in accordance with recent results in the literature that show positive correlation between non-Markovianity and noise-assisted transport efficiencies~\cite{Hauke_PRA_2018,Hauke_PRL_2019,MoreiraPRA2020}.}

\section{Conclusions}\label{Conclu}

We have explored the conditions under which the master
equation describing a driven quantum harmonic oscillator,
interacting with an environment in a non-dissipative way,
is equivalent to the master equation describing 
light propagation in a
dynamically disordered photonic lattice -- the Glauber-Fock 
photonic lattice. One of these conditions is that the noise 
between different sites (waveguides) 
must be correlated. The second condition is to choose a number of waveguides such that the light do not reach the boundary where the sites corresponding to high number states lie.
Further, we have shown that the noise-assisted energy transport 
phenomenon can be observed in this type of systems and
that it is possible to obtain analytical solutions for
certain observables quantities,	e.g. the average photon number. Using these solutions we can readily
predict the maximum amount of energy  transferred 
between all the sites. This, in the Markovian 
scenario, occurs for decoherence rates 
comparable to the energy scale of the system.
For the non-Markovian case, we found that the
range of the dephasing rate, in which the noise-assisted 
transport occurs, is substantially larger.
Our results are in good agreement with recent theoretical~\cite{MoreiraPRA2020,Hauke_PRA_2018}
and experimental~\cite{Hauke_PRL_2019} works showing that non-Markovian environments 
have a strong influence on the energy transport.
Looking forward, and following the ideas of Refs.~\cite{Perez_Leija_2017,de_J_Le_n_Montiel_2019}, 
it would be interesting to go beyond the single excitation regime and derive 
the corresponding master equation of, for example, two correlated particles 
propagating over these stochastic networks affected by non-Markovian noise.

\begin{acknowledgments}
R. R.-A. wants to thanks T.J.G. Apollaro and M. Pezzutto for fruitful \textcolor{black}{initial} discussions. \textcolor{black}{Furthermore, R. R.-A. acknowledges the hospitality of the Max-Born Institute where part of the work was carried out (SPP 1839 Tailored Disorder 2nd Period).} B. \c{C}. is supported by the BAGEP Award of the Science Academy and by the Research Fund of Bah\c{c}e\c{s}ehir University (BAUBAP) under project no: BAP.2019.02.03. \textcolor{black}{R. J. L.-M. thankfully acknowledges financial support by CONACyT under the project CB-2016-01/284372 and by DGAPA-UNAM under the project UNAM-PAPIIT IN102920. A. P.-L. acknowledge partial support by the Deutsche Forschungsgemeinschaft (DFG) within the framework of the DFG priority program 1839 Tailored Disorder.}
\end{acknowledgments}

%\bibliography{references}

\appendix
%\section*{Appendix A}\label{appendixA}
\section{}\label{appendixA}
The time derivative of the density matrix $\sigma_{nm}(t)\equiv\langle \psi_n\psi_m^*\rangle$ is given as
\begin{eqnarray}
\frac{d}{dt}\sigma_{nm}=\Big\langle\psi_m^*\frac{d\psi_n}{dt}+\psi_n\frac{d\psi_m^*}{dt}\Big\rangle,
\end{eqnarray}
where each term can be calculated using the stochastic
Schr\"odinger equation: %of Eq.~(\ref{ssq}):
\begin{subequations} 
\begin{eqnarray}
\psi_m^*\frac{d\psi_n}{dt} =-in\omega\psi_n\psi_m^*-&&in\phi_n(t)\psi_n\psi_m^*
\nonumber\\&&-i\sum_{j\neq n}\kappa_{nj}\psi_j\psi_m^*, \\
\psi_n\frac{d\psi_m^*}{dt} =+im\omega\psi_n\psi_m^*+&&im\phi_m(t)\psi_n\psi_m^*
\nonumber\\&&+i\sum_{j\neq m}\kappa_{jm}\psi_n\psi_j^*.
\end{eqnarray}
\end{subequations}
Performing the stochastic averaging procedure these terms yield
\begin{eqnarray}\label{rhomn_first}
&&\frac{d}{dt}\sigma_{nm}=-i(n\omega-m\omega)\sigma_{nm}
-i\sum_{j\neq n}\kappa_{nj}\sigma_{jm}\nonumber\\ &&
+i{\sum}_{j\neq m}\kappa_{jm}\sigma_{nj}
-in\langle\psi_n\psi_m^*\phi_n(t)\rangle
+im\langle\psi_n\psi_m^*\phi_m(t)\rangle.\nonumber\\
\end{eqnarray}
In particular, the stochastic averages 
in the last two terms of the above equation, one needs to resort to the Novikov's theorem~\cite{novikov1965}:
\begin{subequations}
\begin{eqnarray}
\langle\psi_n\psi_m^*\phi_n(t)\rangle &&=\sum_p\int dt'\langle\phi_n(t)\phi_p(t')\rangle
\left\langle \frac{\delta[\psi_n(t)\psi_m^*(t)]}{\delta\phi_p(t')}\right\rangle,
\nonumber\\ \\
&&=\sum_p\int dt'\Gamma\delta_{np}\delta(t-t')\left\langle\frac{\delta[\psi_n(t)\psi_m^*(t)]}{\delta\phi_p(t')}\right\rangle,
\nonumber\\ \\
&&=\frac{1}{2}\sum_p\Gamma\delta_{np}\left\langle\frac{\delta[\psi_n(t)\psi_m^*(t)]}{\delta\phi_p(t)}\right\rangle,
\label{delta_interpretation}\\%\nonumber\\
&&=\frac{\Gamma}{2}\left\langle\frac{\delta[\psi_n(t)\psi_m^*(t)]}{\delta\phi_n(t)}\right\rangle,
\end{eqnarray}
\end{subequations}
where the operator $\delta/\delta \phi_p(t)$ stands
for the functional derivative with respect to the
stochastic process. Similarly, the second term can be found as
\begin{eqnarray}\label{novikov_second_term}
\langle\psi_n\psi_m^*\phi_m(t)\rangle =\frac{\Gamma}{2}\left\langle\frac{\delta[\psi_n(t)\psi_m^*(t)]}{\delta\phi_m(t)}\right\rangle.
\end{eqnarray}
To obtain Eq.~(\ref{delta_interpretation}) and
Eq.~(\ref{novikov_second_term}) we have used
the fact that, in the Stratonovich interpretation,
$\int\delta(t)=1/2$~\cite{vanKampen1981}.
It is important to mention that the Novikov's theorem 
is only valid for stochastic Gaussian processes, 
that can be both Markovian or non-Markovian as well~\cite{Strunz2004}.
To compute the corresponding functional derivatives 
we need the formal integration of Eq.~(\ref{rhomn_first})
which, before the stochastic average, is
\begin{eqnarray}
\psi_n(t)\psi_m^*(t)=\int_0^t dt'\Big[f&&(\psi_n\psi_m^*,..)
-i n\psi_n\psi_m^*\phi_n(t')
\nonumber\\ &&+im\psi_n\psi_m^*\phi_m(t')\Big],
\end{eqnarray}
where $f(\psi_n\psi_m^*,..)$ represents all the
terms that do not contain stochastic variable, $\phi_n(t)$.
Thus, the functional derivatives are
\begin{subequations}
\begin{eqnarray}
\frac{\delta[\psi_n(t)\psi_m^*(t)]}{\delta\phi_n(t)}=
-i n\psi_n\psi_m^*+i m\psi_n\psi_m^*\delta_{nm},\quad\\
\frac{\delta[\psi_n(t)\psi_m^*(t)]}{\delta\phi_m(t)}=
-i n\psi_n\psi_m^*\delta_{nm}+i m\psi_n\psi_m^*,\quad
\end{eqnarray}
\end{subequations}
in which we have use the identity 
$\delta \phi_p(t')/\delta \phi_q(t)=\delta_{pq}\delta(t'-t)$
~\cite{novikov1965}. 
Using these results we can compute the stochastic 
average for the last 
two terms in Eq.~(\ref{rhomn_first}) as
\begin{subequations}
\begin{eqnarray}
-in\langle\psi_n\psi_m^*\phi_n(t)\rangle &=-\frac{1}{2}\Gamma n^2\rho_{nm}+\frac{1}{2}\Gamma nm\rho_{nm}\delta_{nm},\qquad\\
im\langle\psi_n\psi_m^*\phi_m(t)\rangle &=\frac{1}{2}\Gamma nm\rho_{nm}\delta_{nm}-\frac{1}{2}\Gamma m^2\rho_{nm},\qquad
\end{eqnarray}
\end{subequations}
and as a result we obtain Eq.~(\ref{master_sigma_1}) of the main text.

%\section*{Appendix B}\label{appendixB}
\section{}\label{appendixB}
Applying the Novikov's theorem~\cite{novikov1965} in 
$\langle\psi_n\psi_m^*\Omega_n(t)\rangle$ we get: 
\begin{subequations}
\begin{eqnarray}
\langle\psi_n\psi_m^*\Omega_n(t)\rangle &=\sum_p\int dt'\langle\Omega_n(t)\Omega_p(t')\rangle
\left\langle \frac{\delta[\psi_n(t)\psi_m^*(t)]}{\delta\Omega_p(t')}\right\rangle,
\nonumber\\ \\
&=\sum_p\int dt'\frac{\Gamma\lambda}{2}\delta_{np}e^{-\lambda|t-t'|}\left\langle\frac{\delta[\psi_n(t)\psi_m^*(t)]}{\delta\Omega_p(t')}\right\rangle.
\nonumber\\ \label{noviko_non_markovian}
\end{eqnarray}
\end{subequations}
The final aim of this appendix is to know if the 
master equation of Eq.~(\ref{rhonm_non_first}) 
of the main text, after applying Novikov's theorem in 
Eq.~(\ref{noviko_non_markovian}), will be similar to 
the master equation of Eq.~(\ref{matrix_elements_time}).
Using the formal integration of Eq.~(\ref{rhonm_non_first}),
before doing the stochastic average,
we can compute the functional derivative of 
Eq.~(\ref{noviko_non_markovian}) as follows
%\begin{subequations}
\begin{widetext}
\begin{align}
\frac{\delta[\psi_n(t)\psi_m^*(t)]}{\delta\Omega_p(t')}&=
\frac{\delta}{\delta\Omega_p(t')}
\int_0^tdt''\Big[ f(\psi_n\psi_m^*,..)-in\psi_n\psi_m^*\Omega_n(t'')
+im\psi_n\psi_m^*\Omega_m(t'')\Big],\nonumber\\
&=\int_0^t dt''\Big[-in\psi_n\psi_m^*\frac{\delta\Omega_n(t'')}{\delta\Omega_p(t')}
+im\psi_n\psi_m^*\frac{\delta\Omega_m(t'')}{\delta\Omega_p(t')} \Big],
\nonumber\\
&=\int_0^t dt''\big[-in\psi_n\psi_m^*\delta_{np}\delta(t''-t')
+im\psi_n\psi_m^*\delta_{mp}\delta(t''-t')\big],
\nonumber\\
&=-\frac{i}{2}n\psi_n(t')\psi_m^*(t')\delta_{np}+\frac{i}{2}m\psi_n(t')\psi_m^*(t')\delta_{mp}.
\label{func_deriv}
\end{align}

%\end{subequations}
Performing the stochastic average in 
Eq.~(\ref{func_deriv}) we obtain
\begin{equation}\label{functional_deriv_nom}
\left\langle\frac{\delta[\psi_n(t)\psi_m^*(t')]}{\delta\Omega_p(t')}\right\rangle=
-\frac{i}{2}n\delta_{np}\sigma_{nm}(t')+\frac{i}{2}m\delta_{mp}\sigma_{nm}(t').
\end{equation}
Now we substitute Eq.~(\ref{functional_deriv_nom}) 
in Eq.~(\ref{noviko_non_markovian}):

\begin{subequations}
\begin{align}
\langle\psi_n\psi_m^*\Omega_n(t)\rangle &=
\sum_p\int dt'\frac{\Gamma\lambda}{2}\delta_{np}e^{-\lambda|t-t'|}
\big\{-\frac{i}{2}n\delta_{np}\sigma_{nm}(t')+\frac{i}{2}m\delta_{mp}\sigma_{nm}(t')\big\},
\\
&=-\frac{i}{2}n\int dt'\frac{\Gamma\lambda}{2}e^{-\lambda|t-t'|}\sigma_{nm}(t')
+\frac{i}{2}m\delta_{mn}\int dt'\frac{\Gamma\lambda}{2}e^{-\lambda|t-t'|}\sigma_{nm}(t'),
\\
&\approx-\frac{i}{2}n\sigma_{nm}(t)\int dt'\frac{\Gamma\lambda}{2}e^{-\lambda|t-t'|}
+\frac{i}{2}m\delta_{mn}\sigma_{nm}(t)\int dt'\frac{\Gamma\lambda}{2}e^{-\lambda|t-t'|},\label{slowapprox}
%\\
%&\approx -\frac{i}{2}n\sigma_{nm}(t)\Gamma(t)+\frac{i}{2}m\delta_{mn}\sigma_{nm}(t)\Gamma(t),
\end{align}
\end{subequations}

where we have made an approximation 
$\sigma_{nm}(t')\approx\sigma_{nm}(t)$, i.e., we 
assume that the dynamics of the density matrix 
is slower compared with the dynamics of the
stochastic processes. Under this approximation 
we can perform the integral  of the above equation, %without the term $\sigma_{nm}(t')$, 
\begin{equation}
\int_0^t dt'\frac{\Gamma\lambda}{2}e^{-\lambda|t-t'|}
=\frac{\Gamma}{2}(1-e^{-\lambda t})\equiv\Gamma(t).
\end{equation}
With this result, Eq.~(\ref{slowapprox}) reduces 
to Eq.~(\ref{Psi_nPsi_mOmegt}) of the main text.

\end{widetext}

\end{document}